%% file: main.tex
\documentclass[12pt]{article}

\input{math_commands.tex}

\usepackage{amssymb,amsmath,amsthm,amsfonts,eurosym,geometry,ulem,graphicx,caption,color,setspace,sectsty,comment,footmisc,caption,natbib,pdflscape,subfigure,array,hyperref}
\usepackage{float}
\usepackage{url}

\usepackage{todonotes}

\newtheorem{theorem}{Theorem}

\newtheorem{assumption}[theorem]{Assumption}
\newtheorem{proposition}{Proposition}
\newtheorem{results}[proposition]{Results}

\newcolumntype{L}[1]{>{\raggedright\let\newline\\arraybackslash\hspace{0pt}}m{#1}}
\newcolumntype{C}[1]{>{\centering\let\newline\\arraybackslash\hspace{0pt}}m{#1}}
\newcolumntype{R}[1]{>{\raggedleft\let\newline\\arraybackslash\hspace{0pt}}m{#1}}

\begin{document}

\begin{titlepage}
\title{Crises Do Not Cause Lower Short-Term Growth}%\thanks{abc}}
\author{
Kaiwen Hou\thanks{Columbia Business School, \href{mailto:kaiwen.hou@columbia.edu}{kaiwen.hou@columbia.edu}}
\and David Hou\thanks{Columbia Business School, \href{mailto:yhou24@gsb.columbia.edu}{yhou24@gsb.columbia.edu}}
\and Yang Ouyang\thanks{Columbia Business School, \href{mailto:yo2348@gsb.columbia.edu}{yo2348@gsb.columbia.edu}}
\and Lulu Zhang\thanks{Columbia Business School, \href{mailto:xz3047@gsb.columbia.edu}{xz3047@gsb.columbia.edu}}
\and Aster Liu\thanks{Columbia Business School, \href{mailto:rl3298@gsb.columbia.edu}{rl3298@gsb.columbia.edu}}
}
\date{}
\maketitle
\begin{abstract}
\noindent 
It is commonly believed that financial crises ``lead to" lower growth of a country during the two-year recession period, which can be reflected by their post-crisis GDP growth. However, by contrasting a causal model with a standard prediction one, this paper argues that such a belief is non-causal. To make causal inferences, we design a two-stage staggered difference-in-differences model to estimate the average treatment effects. Interpreting the residuals as the contribution of each crisis to the treatment effects, we astonishingly conclude that cross-sectional crises are often limited to providing relevant causal information to policymakers.
\\
\vspace{0in}
\vspace{0in}

\bigskip
\end{abstract}
\setcounter{page}{0}
\thispagestyle{empty}
\end{titlepage}
\pagebreak \newpage

\singlespacing

\section{Introduction} \label{sec:introduction}
% research question, significance

Financial crises like banking and currency crises have impacted different countries over the past decades. For instance, the global unemployment rate and inflation rate rose significantly after a major financial crisis~\citep{dinccer2018analyzing}; financial crisis weakens stock market performance in several countries~\citep{ahmad2016financial}; the 2008 crisis ``drove down equity levels across the globe, most equity indices were at their 60\% of less of their end of 2006 levels”~\citep{bartram2009no}; and the massive US states bank failures after the 2018 financial crisis~\citep{li2022does}. These critical economic problems above affect countries’ global performance and development. To solve these problems or help the economy to achieve a faster recovery, it is essential for policymakers to examine the causal effects of crises on GDP growth.

There is great heterogeneity in the influence of crises. Some countries have had a much faster recovery following the crises, demonstrated by the higher rate of post-crisis GDP growth, while others have experienced a more prolonged impact as the post-crisis GDP growth has been lower. Therefore, we want to estimate the average treatment effect of recession on the short-term post-crises GDP growth across countries. The significance of this research is to investigate that, more generally, countries are more resilient or vulnerable towards recessions, demonstrated by their post-crises GDP growth. Therefore, we can use this model to predict the future short-term post-crisis GDP growth and provide useful references for policymakers. 

% \al{Why are causal effects of crises in GDP growth significant?}
%From the observation of past financial crises, countries’ economic growth was negatively affected in many areas.  
Previous work~\citep{duffie2019prone} has attempted to make causal arguments about crises, albeit they tend to through a subjective storytelling narrative without rigorous models in causal inference; those who realized the difference~\citep{tiffin2019machine} usually failed to differentiate the causality from correlation. 
%\al{What causal question?  }
Through comparing the results of a prediction model with a staggered difference-in-differences causal model, this paper attempts to find whether there is a causality between financial crises and countries’ short-term GDP growth, and offers insights into the different conclusions drawn from the digressive results of these models.
% \kh{add citation: no statistical method but reached subjective causal conclusion}
Section~\ref{sec:data} introduces the response, treatment, and confounding variables, as well as the years of data we used for analysis. 
%Also, we discussed the handling of missing values, as many distant data are unavailable. 
Section~\ref{sec:LR} discusses the initial model we used for analyzing the linear relationships between the response variable and treatment and confounding variables. This model helps provide insights into how the treatment variable affects the response variable and selects some significant confounding variables through LASSO (least absolute shrinkage and selection operator). However, we believe this model is insufficient to explain the recession's causal effects, so we further improved this model using the Difference-in-Differences method in Section~\ref{sec:DID}. We design a two-stage difference-in-differences regression to obtain the average causal effects of the crises and discover the contributions of different crises to the causal effects. Surprisingly, we conclude that most cross-sectional crises (such as the 2008 financial crisis) do not contribute to the causal effects.\footnote{The codes and results for this paper are available at \url{https://github.com/Davidhyt/causalCrises}.}

%\section{Literature Review} \label{sec:literature}

\section{Data} \label{sec:data}
We collect data from 114 countries after the postwar period (from 1950 to 2022) from various databases of IMF, including Commodity Terms of Trade, Consumer Price Index, International Financial Statistics, International Reserves, and Foreign Currency Liquidity, Financial Development Index, World Revenue Longitudinal Data, and World Economic Outlook Database.\footnote{The databases can be accessed at \url{https://data.imf.org} and \url{https://www.imf.org/en/Publications/SPROLLS/world-economic-outlook-databases}.} The data includes GDP and 20 other variables for each country in each year. After removing and filling in missing values with past data, we have 8,292 effective observations of country-years in total.

% confounding variables/covariates
Serving as confounding variables in studying the causal effects of the crisis on short-term GDP growth, the 20 variables have been selected to cover a broad range of potential confounding variables across different sectors (financial, external, fiscal, and real sectors), outlined in Table~\ref{tab:summaryStats}. Factors like imports and exports of goods and services and foreign currency reserves are classified as external variables. Financial factors including money market interest rates reflect a country's financial status. Real variables, for instance, includes CPI and unemployment rate. We've endeavored to ensure a selection of confounding variables to achieve the widest possible coverage of different aspects of an economy, given that some factors have a short time range of history while others are published in a limited number of countries. A wide range of confounding variables ensures that we eliminate these confounding factors' impacts to quantify better the impact of independent factors, which are crisis events here, on dependent factors.

\begin{figure}[H]
 \centering
 \includegraphics[width=1.3\textwidth]{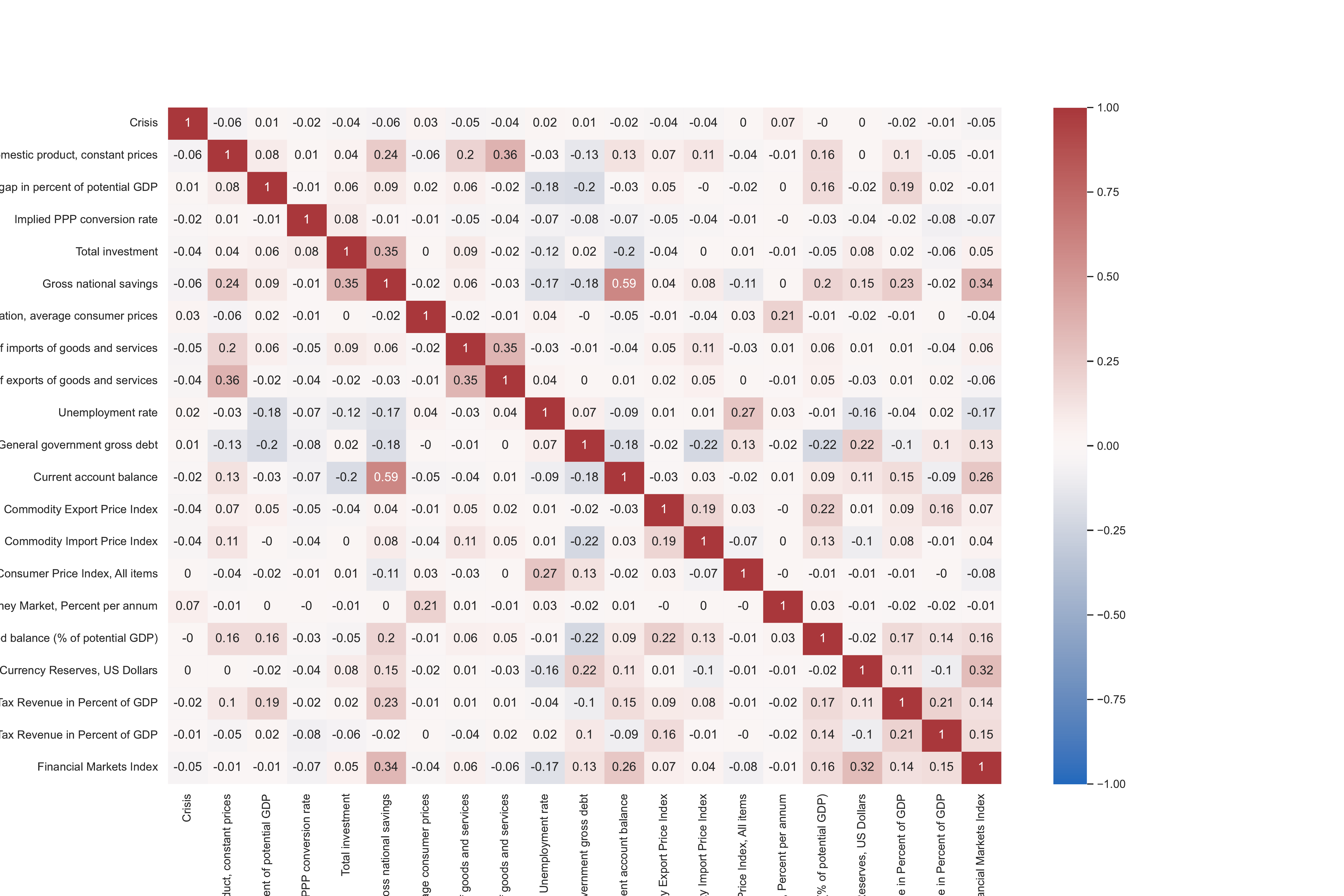}
 \caption{Sample correlation matrix of all selected regressors (including the treatment variable and confounding variables), not showing strong evidence for colinearity.}
 \label{fig:corr}
\end{figure}

The confounding variables are also chosen to ensure the absence of perfect collinearity. Figure~\ref{fig:corr} illustrates the correlation among the regressors, and the darker color means that the correlation is higher in magnitude. For example, current account balance and gross national savings correlate at 0.59. To enhance the robustness of our models, we scale all confounding variables to be within 0 and 1.

\begin{figure}[H]
 \centering
 \includegraphics[width=1.3\textwidth]{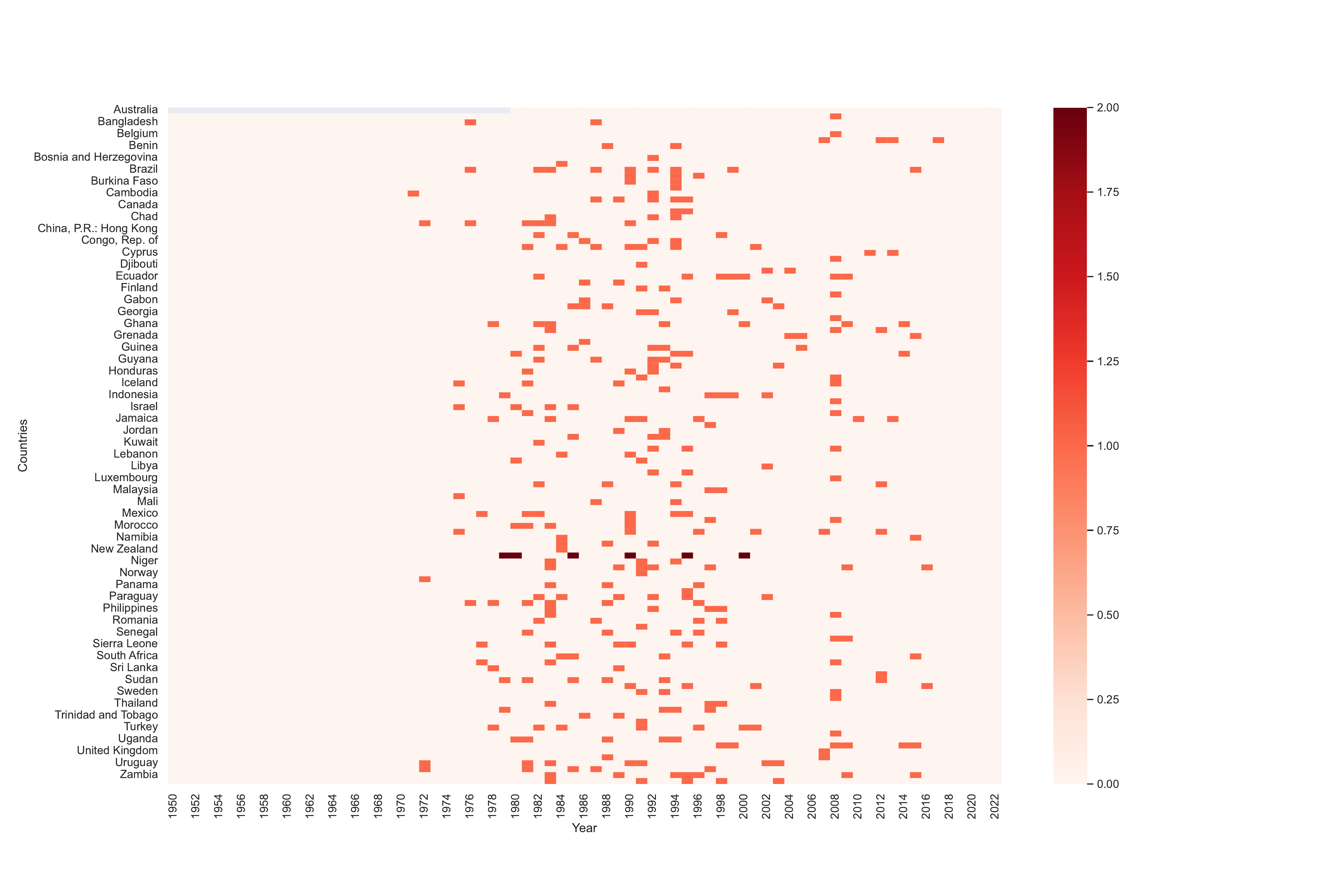}
 \caption{Observed crises in history. Crises for Nicaragua are emphasized with darker colors due to its strong seasonal trend, whose contribution will be explained in Section~\ref{sec:DID}.}
 \label{fig:crises}
\end{figure}

% treatment variable
The treatment variable for crises is obtained based on the~\citet{laeven2018systemic} dataset. Note that here we consider not only banking crises but currency, sovereign, and restructuring ones as well, to more reliably improve the proportion of crises (4.04\% among all observations) in the entire dataset than classical over-sampling or under-sampling methods, (e.g., SMOTE as in~\citet{tiffin2019machine}). All observed crises in history are highlighted in Figure~\ref{fig:crises}.

%The summary statistics of the confounding variables are. 

\section{Multivariate Linear Regression} \label{sec:LR}
% response variable
%\al{need extra \textbf{academic} reference for economic downturn lasting for about 2 years. eg. }
Previous works such as~\citet{demirgucc2006inside} or Table 1 in~\citet{zaman2015economic} generally observe that two years is a typical time for economics to recover after a crisis\footnote{This perception is also common in the industry, such as 17.5 months in \url{https://www.forbes.com/sites/cameronkeng/2018/10/23/recession-is-overdue-by-4-5-years-heres-how-to-prepare/}, or 15-18 months in \url{https://www.acorns.com/learn/investing/how-long-do-economic-downturns-last/}.}, so we use the forward-looking two-year GDP growth rate 
\begin{equation*}
    Y_{it} = \frac{GDP_{i, t+2}}{GDP_{i, t}} - 1
\end{equation*}
for country $i\in\{1\cdots I\}$ at year $t\in\{1\cdots T\}$ to represent the short-term GDP growth rate.
We construct the treatment variable
$$
D_{it} = \begin{cases} 
1 & \text{if a crisis was observed for $i$ at $t$} \\
0 & \text{otherwise}\end{cases}.
$$
Let $\mX_{it}$ denote an observation vector of all confounding variables, including a constant one. These confounding variables are included in the regression to reduce omitted variable bias as far as possible.
Under standard assumptions, consider a simple multivariate regression model
\begin{equation}\label{eq:MLR}
    Y_{it} = \tau D_{it} + \mBeta'\mX_{it} + u_{it}.
\end{equation}

% \begin{equation*}
%     \tau, \mBeta = \argmin 
% \end{equation*}

% Having dealt with the data, our first attempt to find out the influence of crisis on entities' growth rate is trying to fit the data in an ordinary linear regression model.

% In this regression model, two-year GDP growth rate is set as the outcome variable: the aim of this paper is to estimate the likely cost of a crisis, and the two-year growth rate is a good indicator of how the economic entity is affected by it. The existence of crisis is the treatment. As a dummy variable, it takes value 0 when there is no crisis in this year in this country and 1 otherwise. 
 %(to be specified?) 
% It consists of 20 confounding variables listed in Section~\ref{sec:data}.

Since the number of confounding variables can be potentially large (and not small in this case), we apply regularization techniques to prevent overfitting and attain more reliable estimates of the model parameters ($\tau$ and $\mBeta$ in Eq.~\eqref{eq:MLR}). With an additional inductive bias on sparsity~\citep{hou2022spectral}, we also apply LASSO  to fit the regression coefficients, which gives consistent parameter estimates under some regularity conditions~\citep{zhao2006model} or with some adjusted causal models~\citep{athey2018approximate,belloni2014inference}.
% \yo{reference for LASSO being consistent estimators?}
%\kh{consistency~; \url{https://www.r-bloggers.com/2020/09/lasso-and-the-methods-of-causality/}}

%explain the result 
All coefficients $\mBeta$ in this model could be interpreted as the change in GDP growth rate when the corresponding explanatory variables increase by one unit holding other variables constant. The coefficient $\tau$ satisfies
% \kh{TODO: derive $\tau$}
\begin{equation*}
    \tau = \E(Y_{it}\mid \mX_{it},D_{it}=1) - \E(Y_{it}\mid \mX_{it},D_{it}=0),
\end{equation*}
which is the difference of prediction of $Y_{it}$ under treatments or not.

The results are demonstrated in \textbf{Results~\ref{res:MLR}}, where the regression coefficients in $\hat{Y_{it}}^{MLR}$ are estimated without LASSO, only keeping the statistically significant ones, and those in $\hat{Y_{it}}^{MLR,LASSO}$ are shrinkage estimators.
Fixing other regressors, an existence of crisis would lead to a 0.12\% decline in short-term post-crisis GDP growth rate, and the $t$-statistics is -2.378. Also, we can see that inflation influences greatly because fixing other regressors, a 1\% increase in inflation will increase the short-term post-crisis GDP growth rate by 0.64\% and 0.59\%, respectively in the above two models, with the $t$-statistics being 108.424 in the regression model without LASSO. The model has an $R^2$ and adjusted-$R^2$ of 58.9\% and 58.8\%, respectively, which shows that the regressors can explain the response variable quite well. In \textbf{Results~\ref{res:MLR}}, the regressions select variables such as GDP, Inflation, CPI and Interest Rates, which justifies here as they are key macroeconomic indicators that affect GDP growth rate. In addition, comparing the two models, the model without LASSO has a unique factor GDP while the model with LASSO has factors such as Commodity Export Price and Financial Markets Index. The slight difference in models is justified because the main regressors, such as CPI and interest rates, are still the same. 

Remarkably, this regression model can be used for predicting the effects of the recession on short-term post-crisis GDP growth rate through the coefficients. Also, it provides insights into which variables contribute most to the growth rate in terms of prediction.
%\dvh{1. add $R^2$ and adjusted-$R^2$ interpretation; 2. the variables selected by regressions are not insane; 3. w/ w/o LASSO: selected variables are slightly different, but should still make sense. why?, 4. can be used to predict.}
However, one cannot say much about the causal relationship from Eq.~\eqref{eq:MLR} due to several problems. 
% Thus, omitted variable bias always exists in this setting. 
% Although this regression does not explain the causal effects accurately, 
%\kh{add: why this linear regression is not proper for the causal research question}
Due to the limitation of the ordinary multivariate linear regression model, other models are considered to control the omitted variable bias and better analyze the causal relationship.\footnote{Another limitation is having assumed a linear structure to model average treatment effects. There is a whole bunch of literature to tackle nonlinearity, such as~\citet{chernozhukov2018double,farrell2021deep,van2011targeted}.} Among three common regressions, difference-in-differences (DID) regression is chosen, and the reasoning of why this could work is further discussed in Section~\ref{sec:DID}. As to the two-stage least squares (2SLS) method, the setting requires identifying endogenous variables and exogenous variables and then finding corresponding instrument variables to replace endogenous ones. For this treatment, crisis, there is no existing instrument, and manually working out a valid instrument variable would require more rigorous analysis. In terms of regression discontinuity, it is not used in view of the fact that the existence of a crisis is not a treatment that would always occur above a cutoff of a certain value. That is, there is no way to find a cutoff that makes crisis exist continuously above it.

\section{Difference-in-Differences} \label{sec:DID}
% \kh{TODO: define notations Y(0), Y(1), absorbing states}
The nature of our data structure determines our model to be multi-period, and so are the potential outcomes. To generalize the classical analysis of treatment effects to a multi-period setting, denote by $\vzero_t$ the $t$-dimensional vector of zeros, and $\ve_t = (\vzero'_{t-1},1,\vzero'_{T-t})'$. Relaxing the assumption that the treatment $D_{it}$ is an absorbing state as in common staggered adoption design~\citep{roth2022s,athey2022design}, the potential outcomes for country $i$ at $t$ is then represented by $Y_{it}(\vzero_T)$ and $Y_{it}(\sum\limits_{s\in\sS}\ve_s)$ for some set of time indices~$\sS\subset\{1\cdots T\}$. 

As suggested by Section~\ref{sec:data}, the proportion of missing values and the earliest and latest observations are hugely different country-wise, so we are dealing with the imbalanced panel data. Although \citet{de2020two} has assumed a balanced panel of groups, we argue that such an assumption could be relaxed for imbalanced panels to obtain similar theories of DID so that our results remain valid.

Despite some globally impactful crises, such as the 2008 financial crisis, one key challenge for our data is that the time of crises varies across the countries, so the estimate of causal effects is not easily expressed as a difference-in-differences in the basic model. Hence to apply a staggered model (where the treatments are not necessarily imposed at the same time for each $i$, ~\citet{roth2022s}) that also considers the confounding variables $\mX_{it}$, we impose the following \textbf{Assumptions~\ref{ass:ind} and \ref{ass:para}}. 

\begin{assumption}[Independence]\label{ass:ind}
$Y_{it}(\vzero_T), Y_{it}(\sum\limits_{s\in\sS}\ve_s), D_{it}$ and $\mX_{it}$ are jointly independent.
\end{assumption}
The joint independence needed in a staggered model is related yet much stronger than unconfoundedness~\citep{rosenbaum1983central}, which merely requires $D_it$ to be as good as random conditioned on $\mX_{it}$.

% \begin{assumption}[Mean Independence, \citet{wooldridge2010econometric}]\label{ass:meanInd}
% \begin{equation*}
%     % \E\left[ Y_{it}(\vzero_T)-Y_{i,t-1}(\vzero_T)\mid X_{is}, D_{is} \ \forall s\right] =
%     % \E\left[Y_{it}(\vzero_T)-Y_{i,t-1}(\vzero_T)\right].
%     \E\left[ (1-\mL)Y_{it}(\vzero_T) \mid \mX_{i:}, \mD_{i:} \right] =
%     \E\left[(1-\mL)Y_{it}(\vzero_T) \right],
% \end{equation*}
% where $\mL$ is the lag operator.
% \end{assumption}

\begin{assumption}[Conditional Parallel Trends and Mean Independence]\label{ass:para}
\begin{equation*}
    \forall i, \sS: \E\left[ 
    (1-\mL)Y_{it}(\sum\limits_{s\in\sS}\ve_s) \mid \mX_{it}
    \right] =
    \E\left[ 
    (1-\mL)Y_{it}(\vzero_T ) \mid \mX_{it}
    \right] =
    \E\left[(1-\mL)Y_{it}(\vzero_T) \right],
\end{equation*}
where $\mL$ is the lag operator.
\end{assumption}
Mean independence assumes the strong exogeneity condition of the difference $(1-\mL)Y_{it}(\vzero_T)$\footnote{Without loss of generality, this difference has lag 1 in time. Otherwise, such as in our case where $Y_{it}$ looks forward for 2 years, one can always rearrange the data, for instance, by constructing the difference $(1-\Tilde{\mL})Y_{it}(\vzero_T) = (1-\mL^3)Y_{it}(\vzero_T)$, to ensure the difference sensibly formulates the effect of treatment and to avoid look-ahead bias.} and rules out the dip in~\citet{ashenfelter1978estimating}. An alternative is the Assumption S4 of~\citet{de2020two}.
The parallel-trend assumption is crucial to express 
the causal effect $\tau$ as a difference in differences. Formally $\tau$ is a direct generalization of the average treatment effect from~\citet{rosenbaum1983central} as seen from the following theorem.
% \kh{need proof; under \textbf{Assumption~\ref{ass:para}}}
%\begin{theorem}
%\end{theorem}
%\begin{proof}
Notice that under \textbf{Assumption~\ref{ass:para}}, the average treatment effect\footnote{We have implicitly considered the \textit{local} treatment effect at a specific time, but everything is similarly well-defined when one is interested in the effect of a \textit{treatment path} $\mD_{i:}=\sum\limits_{s\in\sS}\ve_s$, although these different paths could be re-parameterized as multi-valued~\citep{de2020several} or continuous treatments~\citep{callaway2021difference}.}
\begin{eqnarray*}
    \tau &=& \E\left[ Y_{it}(\ve_t) - Y_{it}(\vzero_T) \right] \\
    &=& \E\left[ \E(Y_{it}(\ve_t)\mid \mX_{it}) 
    - \E(Y_{it}(\vzero_T)\mid \mX_{it}) \right] \\
    &=& \E\left[ \E(Y_{it}\mid \mX_{it}, \mD_{i:}=\ve_t) 
    - \E(Y_{it}\mid \mX_{it}, \mD_{i:}=\vzero_T) \right] \\
    &=& \E\left[ \E(\mL^{\Delta} Y_{it}\mid \mX_{it}, \mD_{i:}=\ve_t)
    - \E(\mL^{\Delta} Y_{it}\mid \mX_{it}, \mD_{i:}=\vzero_T) \right]
\end{eqnarray*}
%\end{proof}
is simply the expectation of the common trend differences at any time (since the time period~$\Delta$ is arbitrary) and is independent of $i$ and $t$.  Then it is justified to consider the regression specification of DID in Eq.~\eqref{eq:DID}, where $\gamma_i$ is the country fixed effects and $\gamma_t$ is the year fixed effects:
\begin{equation}\label{eq:DID}
    Y_{it} = \gamma_i + \gamma_t + \tau D_{it} + \mBeta'\mX_{it} + u_{it}.
\end{equation}

The DID model has an $R^2$ and adjusted-$R^2$ of 60.3\% and 59.3\%, respectively, which shows that the regressors can explain the response variable quite well. The fitted results (with and without LASSO) are given in \textbf{Results~\ref{res:DID}}, and there is a big difference in regressors after LASSO. We can see that Nicaragua is the only country whose fixed effect is included in the model with LASSO, and one reason can be that crises in Nicaragua happened every five years from 1979/80 to 2000, which illustrates seasonality, and this might be potentially linked to GDP growth rate cycle\footnote{
%\al{reference for business cycles $\sim$ 4.5 years}
The observed seasonality in Nicaragua occurring every five years probably reflects the seasonality in GDP growth rate since approximately 4.5 years is an empirical business cycle, e.g.,~\citet{reinhart2009aftermath,watson1992business}.}. Also, the only common factor between the DID models (with and without LASSO) is the Interest Rate, which justifies intuitively because the interest rate is a critical factor to directly affects demands and GDP growth rate. 

%\dvh{add interpretation to ensure the validity of results, eg. what is unique about Nicaragua among all countries? (crises seem to have seasonality, and what happened to it in history?) Such seasonality occurring every 5 years probably reflects the seasonality in GDP growth rate since 4.5 years is an empirical business cycle (look for some reference for that)}
% \dvh{1. add $R^2$ and adjusted-$R^2$ interpretation; 2. the variables selected by regressions are not insane; 3. w/ w/o LASSO: selected variables are slightly different but should still make sense. why?}

Despite the aforementioned huge difference, $\tau$ is insignificant\footnote{Admittedly, $\tau$ is a lower bound of the real causal effects if there is omitted variable bias since an omitted variable has different signs of correlation with the treatment and response according to \textbf{Results~\ref{res:MLR}}.} no matter whether LASSO is used. To understand this \textit{anti-intuitive} result, we analyze the contribution of each treatment $D_{it}$ on the causal effects $\tau$, we consider the following regression~\citep{roth2022s}
\begin{equation*}
    D_{it} = \alpha + \lambda_i + \lambda_t + \varepsilon_{it},
\end{equation*}
where $\lambda_i$ is the country fixed effects and $\lambda_t$ is the year fixed effects.
\begin{theorem}
$$
    \tau = \E\left[
    \sum\limits_{i,t} \frac{\omega_{it}}{\sum\limits_{j,s}D_{js}}
    \left( Y_{it}(\ve_t) - Y_{it}(\vzero_T) \right)
    \mid D_{it}=1
    \right],
$$
with weights
$$
\omega_{it} = \frac{\varepsilon_{it}\sum\limits_{j,s}D_{js}}
{\sum\limits_{j,s}\varepsilon_{js}D_{js}}.
$$
\end{theorem}

\begin{figure}[H]
 \centering
 \includegraphics[width=1.3\textwidth]{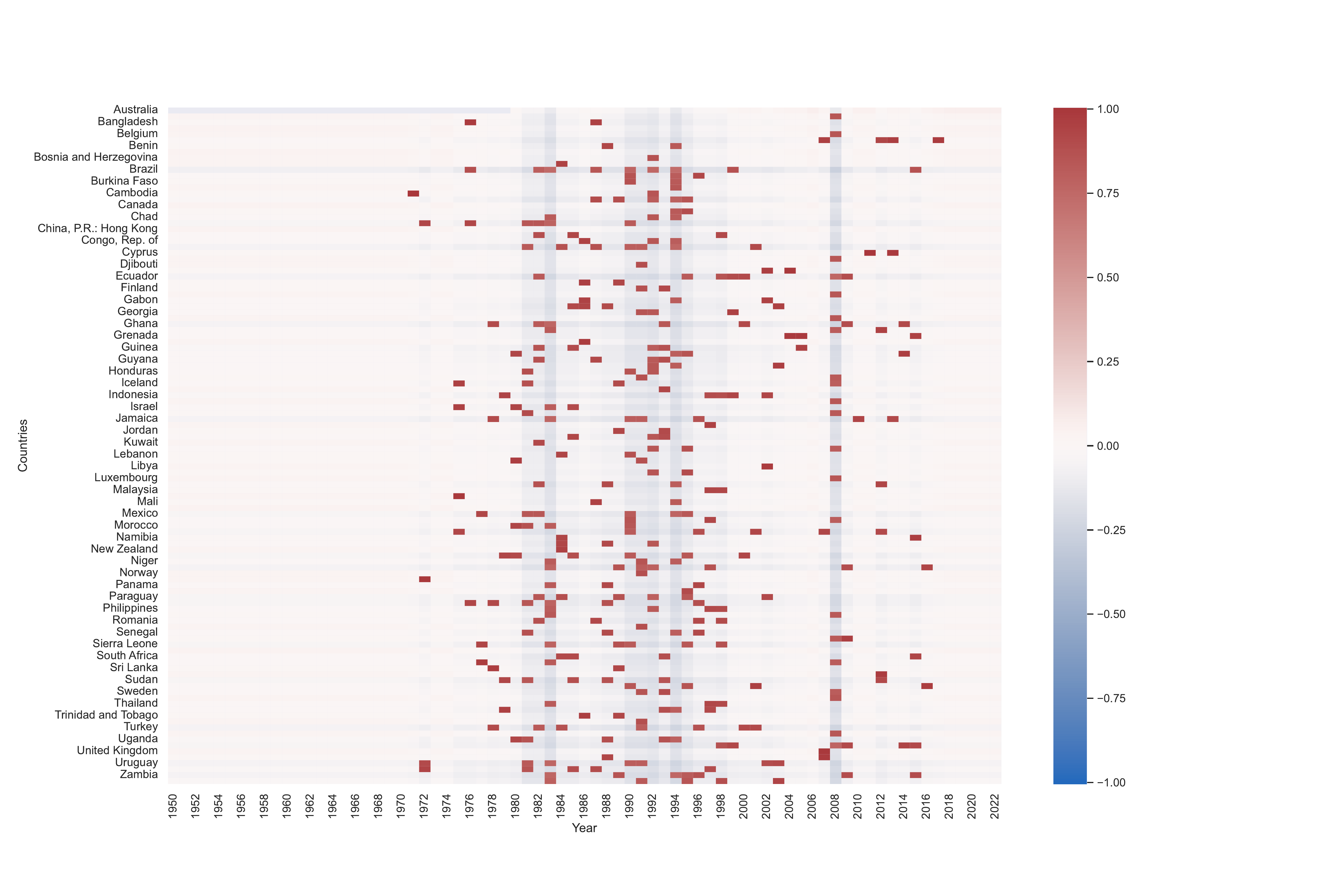}
 \caption{Contribution of each country in each year to the causal effects $\tau$}
 \label{fig:countries}
\end{figure}

Figure~\ref{fig:countries} illustrates each country's contribution each year to the causal effects, and the darker color means that the contribution is higher. For example, the years 2008 and 1994 have a darker color than many other years because of the financial crisis and the Fed hike period. The red color represents a positive residual (hence positive weights of contribution to $\tau$), and the blue color represents a negative residual. 

To our surprise, we observe that blue vertical strips correspond to negative weights, which often match the cross-sectional crisis periods, such as in 1983, 1994, and 2008, indicating that they negatively contribute to the causal effect of growth reduction. On the other hand, the scattered crises in red mostly contribute to the causal effect, which is largely heterogeneous across both countries and time. Thus, we argue that the former is less relevant for policymakers to tackle crises than the latter.

\section{Conclusion} \label{sec:conclusion}

% \al{implications to policy-making: why do policy makers need causal information of crises? eg. estimate likely cost of a crisis, and so what?}
This research paper demonstrates that there is no casual effect between financial crises and countries’ GDP growth, but the correlation effect does exist. This result will benefit policymakers, who must respond relatively quickly to face financial crises. For example, some policymakers use monetary policy or fiscal policy to recover the economy in the short run. However, these policies might lead to negative effects in the long run. Policymakers must take lessons from crisis experiences to better enact policies after crises. Knowing that there is only a correlation effect between financial crises and countries’ GDP growth, policymakers can shift their focus towards factors that directly influence economic growth \textit{if there are any at all}. %They can then examine the causal effect between specific economic problems, such as increasing unemployment rate and inflation rate, and GDP growth.

% \dvh{maybe we \textit{can} predict, but it's just not causal? -- from Kaiwen}
Note that Section~\ref{sec:DID} alone does not imply the absence of a causal relationship between the treatment and the response. In the initial model, we have seen the crisis coefficient be significant, but in our subsequent model to analyze causal effects, the coefficient of crisis has turned insignificant. With the two results together, we can conclude that all the causal effects are explained by some confounding variables, where crises do not play a role. Therefore, policymakers can use the crisis to predict the post-crisis short-term GDP growth rate, but they can't use the model to derive any causal effects of crises.

% \kh{future studies}
Due to the continuous nature of confounding variables, we have not yet come up with a hypothesis test for conditional parallel trends. To make the theory more statistically rigorous, further studies can involve either generalizing the Sniff tests~\citep{roth2022pretest} in the confounding context or translating the continuous confounding variables to discrete ones.
%adding more confounding variables so that the treatment will seem almost random conditioned on them. 

\singlespacing
\setlength\bibsep{0pt}
\bibliographystyle{aea}
\bibliography{literature}

\clearpage

\onehalfspacing

\section*{Tables} \label{sec:tab}
\addcontentsline{toc}{section}{Tables}

\begin{table}[H]
    \centering
    \begin{tabular}{c|c|c|c}
     \hline\hline
     Variable Names &Abbr& Mean & Std \\
     \hline\hline
     Gross domestic product, constant prices& GDP   & 3.29    &5.95\\
     Output gap in percent of potential GDP& Gap &   -0.43  & 2.93  \\
     Implied PPP conversion rate&PPP&   148.72  & 571.10  \\
     Total investment&Invt &23.11 & 9.72\\
     Gross national savings&Sav    &19.42 & 11.93\\
     Inflation, average consumer prices&Inf&   26.90  & 292.66\\
     Volume of imports of goods and services&Imp&   5.47  & 16.89\\
     Volume of exports of goods and services&Exp&   5.64  & 20.60\\
     Unemployment rate&Unemp & 8.61  & 5.68   \\
     General government gross debt&Debt & 60.38  & 44.57\\
     Current account balance&CAB & -2.79  & 9.91\\
     Commodity Export Price Index&CEPI & 95.84  & 9.92\\
     Commodity Import Price Index&CIPI & 95.75  & 4.80\\
     Consumer Price Index, All items&CPI& 149.89 & 691.29\\
     Financial, Interest Rates, Money Market &IR & 25.60  & 464.25\\
     Cyclically adjusted balance (\% of potential GDP)&Adj & -2.65  & 3.67\\
     Official Reserve Assets, Foreign Currency Reserves &FCR & $6.90\times10^{10}$  & $1.66\times10^{11}$\\
     Corporate Income Tax Revenue in Percent of GDP&CITR & 3.00  & 2.25\\
     Goods and Services Tax Revenue in Percent of GDP&Tax & 7.92  & 3.84\\
     Financial Markets Index&FMI & 0.19  & 0.24\\
     \hline\hline
    \end{tabular}
    \caption{Summary statistics of confounding variables}
    \label{tab:summaryStats}
\end{table}

\clearpage

\section*{Figures} \label{sec:fig}
\addcontentsline{toc}{section}{Figures}

%\dvh{make heatmap of $D_{it}$}

\clearpage

\section*{Appendix A. Regression Results} \label{app:regression}
\addcontentsline{toc}{section}{Appendix A}
For complete outputs of regression models, including $t$-statistics, $P$-values, confidence intervals, and other evaluation metrics of regression, refer to  \url{https://github.com/Davidhyt/causalCrises/blob/main/MLR.ipynb} and \url{https://github.com/Davidhyt/causalCrises/blob/main/DID.ipynb}. The following reports the point estimates of \textit{statistically significant} coefficients.
% \yo{add the following variable names to Table 1, otherwise they are meaningless.}
\begin{results}[Multivariate Linear Regression]\label{res:MLR}
\begin{eqnarray*}
    \hat{Y_{it}}^{MLR} &=& -0.0012D_{it} \\
    &+& 0.0128GDP_{it}+0.6437Inf_{it} -0.0034CPI_{it}-0.1728IR_{it}+0.0032CITR_{it} \\
    \hat{Y_{it}}^{MLR,LASSO} &=& -0.0039-0.0012D_{it} \\
    &+& 0.5930Inf_{it}+0.0001CEPI_{it}-0.0021CPI_{it}
    -0.0970IR_{it}+0.0011CITR_{it} \\
    &+& 0.0002FMI_{it}.
\end{eqnarray*}
\end{results}

% \yo{perhaps use full name for countries:  Done}
% \yo{fixed effects no subscripts ``it" $=>$ change to indicators $\mI$:  Done}
% \dvh{!!! $\hat{Y_{it}}^{DID,LASSO}$ can further shrinkage the parameters (currently too many parameters, so we cannot tell if $\tau$ is meaningful; ideally, it should be 0)}
\begin{results}[Difference-in-Differences]\label{res:DID}
\begin{eqnarray*}
    \hat{Y_{it}}^{DID} &=& 0.0136Gap_{it} +0.0024PPP_{it}+0.6456Inf_{it}-0.0085Exp_{it}-0.1655IR_{it} \\ 
    &+& 0.007\mI_{Nicaragua} -0.0087\mI_{Peru}-0.006\mI_{Turkmenistan}-0.004\mI_{Ukraine}\\
    &+& 0.003\mI_{1987} +0.0034\mI_{1988} -0.0023\mI_{1989}-0.004\mI_{1990}-0.0035\mI_{1993} \\
    \hat{Y_{it}}^{DID,LASSO} &=& -0.0014+0.2359Inf_{it}+0.0086\mI_{Nicaragua}
\end{eqnarray*}
\end{results}

% \clearpage

% \section*{Appendix B. Proofs} \label{app:regression}
% \addcontentsline{toc}{section}{Appendix B}

\end{document}

%% file: math_commands.tex
\usepackage{amsmath,amsfonts,bm}

\def\1{\bm{1}}

\def\vzero{{\bm{0}}}

\def\ve{{\bm{e}}}

\def\mD{{\bm{D}}}

\def\mI{{\bm{I}}}

\def\mL{{\bm{L}}}

\def\mX{{\bm{X}}}

\def\mBeta{{\bm{\beta}}}

\DeclareMathAlphabet{\mathsfit}{\encodingdefault}{\sfdefault}{m}{sl}
\SetMathAlphabet{\mathsfit}{bold}{\encodingdefault}{\sfdefault}{bx}{n}

\def\sS{{\mathbb{S}}}

\newcommand{\E}{\mathbb{E}}